# Structural and electronic phase evolution of Tin dioxide

Sudipta Mahana[1], Pitamber Sapkota[2], Saptarshi Ghosh[3], U. Manju[4], D. Topwal[1],*

[1]*Institute of physics, Sachivalaya Marg, Bhubaneswar - 751005, India*
[2]*Department of Physics, Central Campus, Tribhuvan University Kirtipur, Kathmandu, Nepal*
[3]*Sensor and Actuator Division, CSIR-Central Glass and Ceramic Research Institute,
196, Raja S. C. Mullick Road, Kolkata - 700032, India and*
[4]*Materials Characterisation Division, CSIR-Central Glass and Ceramic Research Institute,
196, Raja S. C. Mullick Road, Kolkata - 700032, India**

(Dated: June 27, 2016)

We investigate the effect of controlled annealing on the structural and electronic phase evolution of Tin dioxide from Tin (II) oxyhydroxide prepared by simple precipitation method. Thermogravimetric analysis suggests a complex weight loss-gain process involved, passing through an intermediate phase of tin oxide nanoparticles. The probable structural and electronic phase evolution is discussed using detailed X-ray diffraction and X-ray photoelectron spectroscopy investigations.

PACS numbers:

## I. INTRODUCTION

Tin dioxide ($SnO_2$) is a known n-type wide band-gap semiconductor. Due to its outstanding electrical, optical and electrochemical properties it is proven to be an exciting candidate for basic sciences research as well as for technological applications. Extensive studies have been conducted on this material to explore its potential application as catalysts[1], solar cells[2], optoelectronic devices[3], gas sensors[4, 5], transparent electrodes/conductors[6, 7] etc. $SnO_2$ is widely used as a base material for solid state gas sensors in alarms for sensing alcohol and moisture[8, 9]. $SnO_2$ powders and films for the above mentioned applications can be produced by various synthesis methods, such as hydrothermal synthesis[8], sonochemical[10], sol-gel[11], spray-pyrolysis[12], chemical vapour deposition(CVD)[13] and wet chemical method[14].

Chemical precipitation method is one of the most widely used techniques to produce $SnO_2$ powders because it is easier to implement, is cost effective and could be carried out at ambient conditions. One of the most commonly used tin precursor is $SnCl_4 \cdot 5H_2O$[15, 16] where Sn is already in tetravalent oxidation state in the starting material. However, some studies have also been done by using $SnCl_2 \cdot 2H_2O$ as starting material where the initial divalent oxidation state of Sn is transformed to tetravalent state by annealing the sample in air at relatively higher temperatures, to form pure $SnO_2$. Alternatively, if synthesis is carried out through sonication-assisted precipitation technique then depending on the ultrasonic power various phases of tin oxide, namely, SnO or $SnO_2$ can be directly obtained even at room temperature[17].

In the present work, we discuss the sample growth through a simple chemical precipitation route using $SnCl_2 \cdot 2H_2O$ as the precursor and its transformation into $SnO_2$ phase. Thermogravimetric Analysis (TGA) analysis showed a complex weight loss and gain behaviour during the entire heating process. Hence we annealed the sample at some selected points along the TGA plot and performed detailed X-ray diffraction (XRD) studies together with Rietveld refinement to quantitatively analyze the structural phase evolution during the growth process. Over the years, X-ray photoelectron spectroscopy (XPS) technique has established its unique capability in probing the elemental and chemical state of materials. Hence utilising this capability, we performed detailed core level and valence band photoemission measurements thereby helping in understanding the underlying electronic phase evolution during the growth process.

## II. EXPERIMENTAL

A clear solution containing $Sn^{2+}$ (0.17 M) was prepared by dissolving stannous chloride dihydrate ($SnCl_2 \cdot 2H_2O$) in deionized water under continuous stirring. To increase the solubility, solution was mildly heated and a few drops of conc. HCl was added with continuous stirring. The resulting solution had a pH of 1. Subsequently ammonia solution was added in a drop wise manner in the obtained stannous chloride solution after bringing it to room temperature. Precipitation occurred as the pH of the solution gradually increased to 8. White precipitate obtained during the process was filtered out after repeatedly washing it with deionized water to remove chloride ions, excess $NH_4OH$ and other byproducts like $NH_4Cl$ etc. Filtered sample was then dried in the refrigerator and subsequently annealed at various temperatures up to 950 $°C$ in air.

Thermogravimetric analysis of the as synthesised sample (obtained after drying in refrigerator) was carried out in the temperature range from room temperature up to 1000 $°C$ at a rate of 7 $°C$/min in a mixture of 80% nitrogen and 20% oxygen flow in STA 499 Netzsch instrument. X-ray diffraction measurements for structural phase analysis of the samples were carried out with Bruker D8 Advance X-ray diffractometer using Cu $K_\alpha$ radiation, while Rietveld refinements for determining the crystal structure and phase composition of the samples were carried out using TOPAS-3 software. X-ray photoelectron spectroscopic investigations were performed on some selected samples to study the electronic phase evolution of the compounds involved. These experiments were carried out in a PHI-5000 Versaprobe II Scanning XPS Microprobe using monochromatic Al $K_\alpha$ radiations at room temperature [18].

---

*Electronic address: dinesh.topwal@iopb.res.in,dinesh.topwal@gmail.com

### III. RESULTS AND DISCUSSION

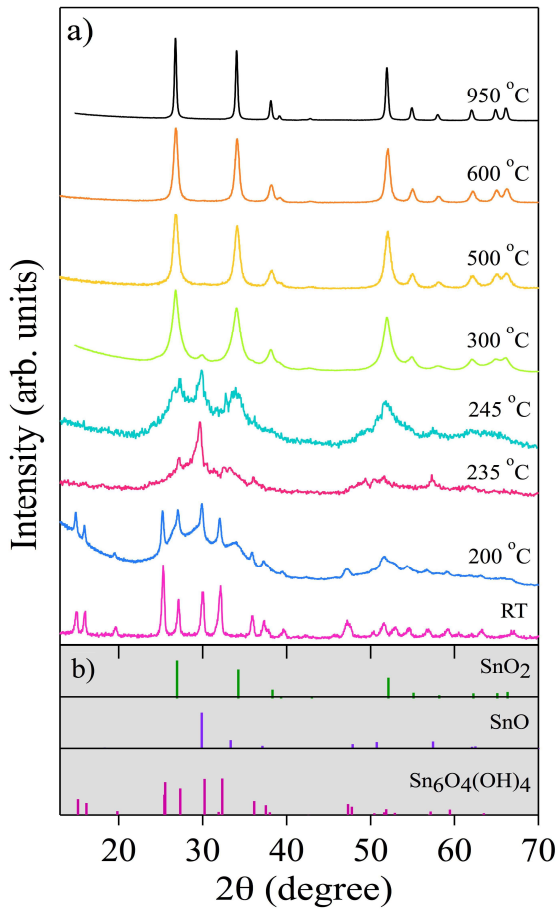

FIG. 1: (a) X-ray diffractogram of the samples annealed at different temperatures as indicated. (b) Reference X-ray diffractogram obtained from the ICDD-PDF data base for various patterns $Sn_6O_4(OH)_4$ (46-1486), SnO (85-0423) and $SnO_2$ (88-0287)

Figure 1(a) shows powder x-ray diffraction patterns of as-prepared and heat-treated samples at different temperatures as indicated in the figure. Diffraction pattern marked RT is that of as prepared sample which was obtained after drying the sample in refrigerator. It suggests formation of a single phase of Tin (II) oxyhydroxide ($Sn_6O_4(OH)_4$) with high degree of crystallinity[19–21]. TGA analysis of this sample is depicted in Figure 2 which shows an initial weight loss while heating, followed by a gain in weight after passing through a minimum at around 235 °C. Beyond 500 °C weight of the sample shows saturation where a very slow decrease in weight could be noted till 1000 °C, the highest temperature of measurement. There are no detailed discussions available in the literature about the crystallographic phase evolution of tin oxide during this complex weight loss and gain process [17, 19]. To obtain a detailed qualitative understanding, we subjected the sample to annealing at different temperatures along the TGA curve, collecting XRD patterns and performing compositional phase analysis to have a better understanding of the process/phases involved.

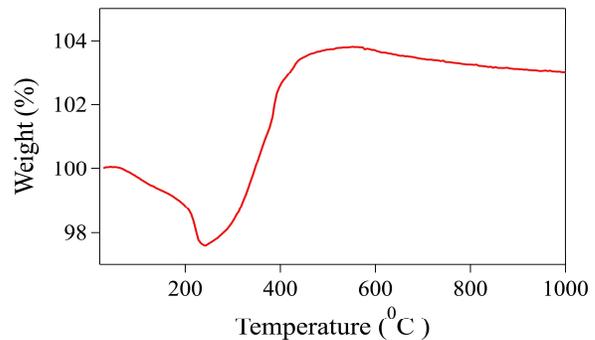

FIG. 2: TGA plot for the $Sn_6O_4(OH)_4$ sample

Initial weight loss of the sample as observed from the TGA plots up to 100 °C and a little beyond can be attributed to the loss of adsorbed water together with dehydration[19] and initiation of the decomposition process of $Sn_6O_4(OH)_4$ to form SnO [17] as suggested in equation (1).

$$Sn_6O_4(OH)_4 \rightarrow 6SnO + 2H_2O \quad (1)$$

During the initial process of loss of adsorbed water no noticeable change in the XRD pattern is observed between the as prepared sample and the sample heated at 100 °C. However, after calcining the sample at 200 °C there is a distinct change in the XRD pattern (see Figure 1(a)). Diffraction pattern in the $2\theta$ value range from 25° to 40° seems to ride on a hump. Also extra peaks appear at $2\theta$ value corresponding to 33.7° and shoulders at 29.2° and 26.4° suggesting the formation of $SnO_2$ and SnO phase as indicated from the standard XRD patterns reported in ICDD-PDF data base plotted in Figure 1(b) for reference. Temperature around 200 °C hence marks the initiation of the decomposition process of $Sn_6O_4(OH)_4$ as indicated in equation (1). However, the formation of $SnO_2$ phase cannot be explained by this equation alone. The presence of $Sn_6O_4(OH)_4$ along with SnO and $SnO_2$ at 200 °C signifies the concurrent occurrence of the processes mentioned in equations (1) and further initiation of oxidation of SnO to form $SnO_2$ as will be discussed later.

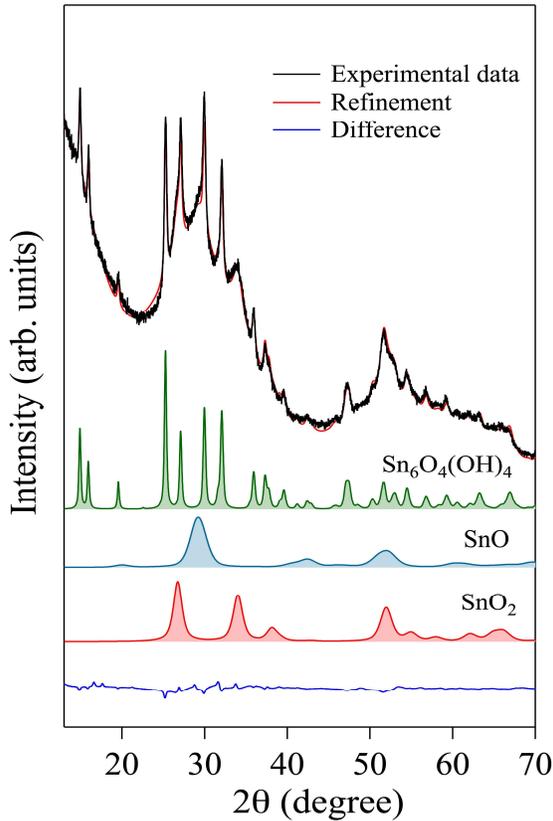

FIG. 3: Rietveld refinement and quantitative phase analysis of the 200°C heat treated sample using TOPAS3 software. Different phases present in the sample are indicated in the plot.

Rietveld refinement and quantitative phase analysis of the 200 °C heat treated sample was carried out and the results are summarized in Figure 3. Best fit to the experimental data was obtained when Rietveld phase analysis gave relative weight fraction of 43.5%, 18% and 38.5% for $Sn_6O_4(OH)_4$, SnO and $SnO_2$, respectively suggesting a mixed phase at this temperature. Also the crystallite size of SnO and $SnO_2$ could be estimated to be around 16 and 10 nm, respectively from Rietveld refinement . These results suggest that after calcining the sample at 200 °C, in an overall matrix of $Sn_6O_4(OH)_4$, nano particles of SnO and $SnO_2$ start growing at certain sites. With a further increase in temperature of few degrees i.e. up to 235 °C, sharp weight loss is observed in the TGA plot and the corresponding XRD pattern shows a complete absence of $Sn_6O_4(OH)_4$ phase and a strong emergence of SnO phase. Rietveld phase analysis of 235 °C heat treated sample gave relative weight fraction of 60% and 40% for SnO and $SnO_2$, respectively for the best fit[22]. It may be noted that the concentration of $SnO_2$ phase remained almost constant with annealing in 200-235 °C temperature range, the concentration of SnO phase rapidly grew at the expense of $Sn_6O_4(OH)_4$ phase resulting in a sharp weight loss as indicated in equation (1). Further heating of the sample in air beyond 235 °C results in the oxidation of SnO to form $SnO_2$ as observed by XRD; this process also results in weight gain as observed from the TGA plots. It is observed that trace amount of SnO survives all the way up to 380 °C. At 500 °C and beyond, XRD pattern confirms the completion of the phase transformation to the pure tetragonal rutile structure of $SnO_2$ without any impurity phase. An insignificant weight loss is observed beyond 500 °C which can be attributed to the presence of trace amount of hydroxide component observed by Fourier Transform Infrared spectroscopy[19] or recovery of some oxygen deficiency in $SnO_2$ which is beyond the detectable limit of XRD. To summarise, there seems to be certain sites in the sea of $Sn_6O_4(OH)_4$ matrix which serve as nucleating centres for the growth of SnO and $SnO_2$ nano particles. These sites are created when temperature is raised beyond 100°C. Further, the number of these sites could be dependent on the sample preparation conditions. With increasing temperature, size of $SnO_2$ nano particles increases slowly as expected. However, the number of $SnO_2$ sites remain almost constant up to 235 °C which could be speculated that the kinetics required for their growth is not satisfied till 235 °C, even though in this temperature region the majority matrix completely transforms from $Sn_6O_4(OH)$ to SnO. Beyond 235 °C $SnO_2$ nano particles grow rapidly at the expense of SnO matrix or according to equation (3). During the growth process crystallite size of $SnO_2$ particles is found to increase gradually from 10 nm, 13 nm, 20 nm, 33 nm, 49 nm, 66 nm and 125 nm for annealing temperatures 200 °C, 235 °C, 245 °C, 300 °C, 500 °C, 600 °C and 950 °C, respectively.

Two different routes have been suggested in literature for the oxidation of the SnO sample to form $SnO_2$ [17, 23]

$2SnO \rightarrow Sn + SnO_2$ (2)
$SnO + 1/2\, O \rightarrow SnO_2$ (3)

However, our TGA results does not support the reaction mentioned in equation (2), according to which after the initial loss in weight of the sample due to the decomposition of $Sn_6O_4(OH)_4$ to form SnO (as mentioned in equation (1)), weight of the sample should remain constant with temperature. Instead, we notice an increase in weight of the sample beyond 235 °C, which can be attributed to oxidation of SnO to form $SnO_2$ as depicted in equation (3). This observation is validated further by our XRD patterns where we do not observe any traces of Sn. We also annealed $Sn_6O_4(OH)_4$ sample in inert atmosphere to deny the reaction of any external Oxygen in the process of annealing. XRD pattern of this sample showed traces of Sn and TGA plot did not show any increase in weight of the sample after the initial weight loss as expected from equation (1) and (2).

Photoemission spectroscopy was employed to study the electronic properties / transition of tin oxide at various stages of annealing. Figure 4 shows Sn $3d$ core level photoemission spectra recorded for three compounds, namely, as prepared sample marked as RT, sample heated at 245 °C and 950 °C. The spin-orbit splitting between Sn $3d_{5/2}$ and Sn $3d_{3/2}$ for all the three compounds is found to be 8.4 eV which is in good agreement with the values reported in the literature [24, 25]. Binding energy position for the Sn $3d_{5/2}$ peak for 950 °C heat treated sample is found to be 486.6 eV, with Full Width Half Maximum (FWHM) of about 1.18 eV. It may here be noted that the FWHM of the Sn $3d$ peaks in our case is much smaller than in many other reports in the literature [24, 26]. This value corresponds to that of $SnO_2$ *i.e.* Sn in $Sn^{4+}$ state [26–29]. Noticeably binding energy, spectral shape and the FWHM of the 245 °C heat treated sample resembles very closely with that of 950 °C sample. From quantitative phase analysis of XRD patterns, the sample annealed at 245 °C is composed of $SnO_2$ and SnO in the ratio 57:43, which should also reflect in the XPS spectra. However, analysis based on the chemical shift of Sn $3d$ spectrum and its deconvolution cannot seem to distinguish between formal valencies of tin in its oxidised form, namely $Sn^{2+}$ in SnO and $Sn^{4+}$ in $SnO_2$ [30]. It is argued that chemical shift due to the change in free-ion potential between $Sn^{2+}$ and $Sn^{4+}$ is cancelled by the change in Madelung potential at tin sites between SnO and $SnO_2$. In contrast Sn $3d$ spectrum of RT ($Sn_6O_4(OH)_4$) appears to be shifted towards lower binding energy by (~0.5 eV) and has a higher FWHM of 1.6 eV.

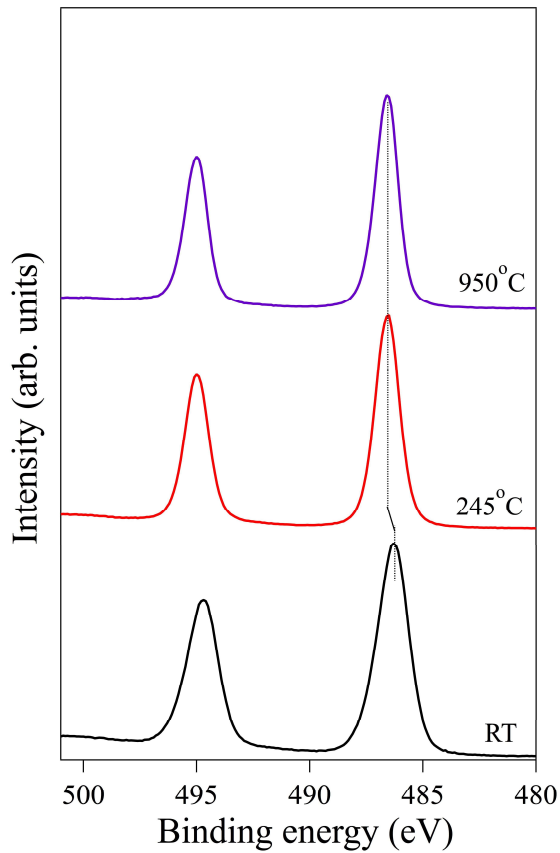

FIG. 4: XPS core level spectra of Sn $3d$ measured using monochromatized Al $K_\alpha$ source for samples treated at room temperature (RT), 245 °C and 950 °C.

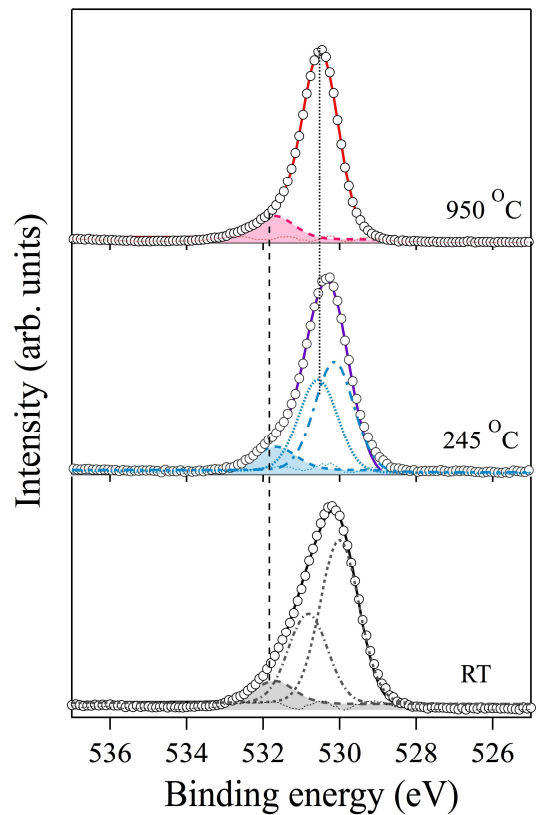

FIG. 5: O $1s$ core level spectra of samples treated at room temperature (RT), 245 °C and 950 °C. Open circles shows the experimental spectra, thick solid line shows the fitted spectra and the thin dotted line shows the difference between the two. Dashed lines correspond to the various deconvoluted spectral components.

Since chemical shifts in Sn $3d$ core levels for SnO and $SnO_2$ appear to the same, it cannot be used as a criteria for distinguishing them. However, O $1s$ and valance band spectra (Figure 5 and Figure 6, respectively) shows distinctively different features and are hence suitable for this purpose. Figure 5 shows O $1s$ spectra from $Sn_6O_4(OH)_4$ sample (marked as RT), 245 °C heat treated sample and $SnO_2$ sample (950 °C heat treated sample). On spectral decomposition, it is noted that O $1s$ spectra of all the three samples have a peak at ~531.7 eV binding energy, with a FWHM of ~1.13 eV and almost equal area in all the three cases, it is marked by the shaded area in the figure for help of view. We believe that this highest binding energy peak corresponds to the chemisorbed species of oxygen [27, 31]. Apart from this O $1s$ spectrum of $SnO_2$ sample could be fitted reasonably well with only one more peak with binding energy of 530.5 eV, this oxygen species corresponds to O-$Sn^{4+}$ [26, 27, 29, 32]. Similarly O $1s$ peak for 245 °C heat treated sample can be fitted with two more components (other than the chemisorbed species) having binding energies of 530.6 eV and 530.1 eV with a FWHM of 1.2 eV and 1.1 eV, respectively. As mentioned above 530.6 eV peak can be assigned to O-$Sn^{4+}$ species while 530.1 eV is found to be associated with O-$Sn^{2+}$ species arising out of SnO as reported in literature [26, 27, 29, 31]. Area under these deconvoluted peaks can also provide information about the phase composition of the sample. Here the ratio between these peaks is obtained to be 56:44 which is in agreement with the results obtained from XRD measurements where relative weight fraction between $SnO_2$ and SnO for the 245 °C heat treated sample is about 57:43, even though techniques like XRD and XPS looks at different depth profiles of the sample; XRD is a bulk measurement where as XPS is a surface sensitive technique thereby providing information relatively from the surface layers. O1s spectra from $Sn_6O_4(OH)_4$ (marked as RT) looks broadest among all the three. However, it is best fitted by three components one chemisorbed species (marked by the shaded area) and the other two components having binding energies of 530 eV and 530.8 eV with FWHM of 1.18 eV and 1.19 eV, respectively. In Tin (II) oxyhydroxide ($Sn_6O_4(OH)_4$) Sn is coordinated by two different types of oxygen species one simple oxide O and other hydroxy O atoms [20] these two different types of oxygen environments give rise to the two components mentioned above. Similar results are observed for other oxyhydroxide compounds like for iron(III) oxyhydroxide (FeOOH), where higher binding energy component corresponds to $(OH)^-$ species and peak at 530.0 eV represent O from oxide nearly invariant to $Fe_2O_3$ [33–37]. By similar arguments, peak at 530 eV represents O of oxide species almost same as in SnO [26, 27, 29, 31] and higher BE peak (530.8 eV) corresponds to O of hydroxide species.

Figure 6 shows the valence band spectra from the corresponding samples discussed above. Comparing the valence band spectra of 950 °C and 245 °C heat treated samples, it is noted that these two spectra are very similar and are mainly dominated with the contributions from $SnO_2$ [24, 27] where peaks about 4.5 eV, 7.5 eV and 10.6 eV binding energies originate from O 2$p$ nonbonding, Sn 5$s$ - O 2$p$ and Sn 5$p$ - O 2$p$ hybridized states, respectively. The only distinct difference in the valence band lies at $\sim$ 2.8 eV binding energy where a shoulder appears at the rising edge towards the lower binding energy in the case of 245 °C heat treated sample. Inset to the Figure 6 shows XPS valence band of SnO adapted from literature[24] and valence band of $SnO_2$ (from 950 °C treated sample). Taking valence band contributions in the ratio 56:44 for $SnO_2$ and SnO, as suggested by the XRD analysis, the resulting generated spectra (after addition) is similar to that of the 245 °C heat treated sample (shown in the inset). Origin of 2.8 eV shoulder can hence be attributed to the presence of SnO in the sample. These are Sn 5$s$ derived levels which in $SnO_2$ forms the bottom of the conduction band, but becomes a part of the valence band in SnO, in order to accommodate additional electrons on tin cation. Valence band spectra of $Sn_6O_4(OH)_4$ is marked as RT in Figure 6, unlike other two spectrum it shows strong spectral features at $\sim$ 2.2 eV along with the other prominent features at 4.5, 7 and 9.5 eV, which are also shifted towards lower binding energy. Here too the 2.2 eV feature may arise from Sn 5$s$ derived states like in SnO as Sn is in divalent oxidation state. Further detailed experimental and theoretical analysis is required to have a better understanding of the electronic structure of this complex.

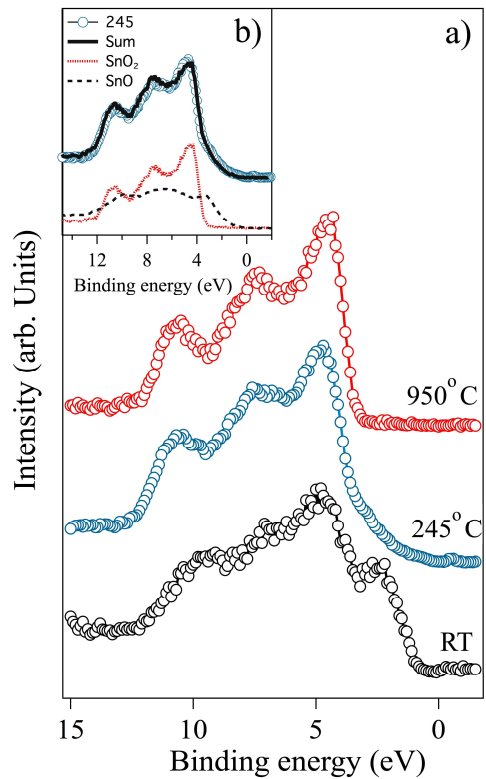

FIG. 6: XPS valence-band spectra of samples treated at room temperature (RT), 245°C and 950°C as indicated in the figure. Inset shows SnO data adapted from ref[24]

## IV. CONCLUSION

In this work we have successfully prepared tin (II) oxyhydroxide sample by precipitation method. The sample is found to be stable up to 100 °C. Using X-ray diffraction and thermogravimetric analysis it is found that beyond 100 °C sample starts decomposing first as nano particles of SnO and $SnO_2$ at certain nucleating sites, though the majority phase of the sample still continues to remains as Tin (II) oxyhydroxide. With further increase in annealing temperature the majority phase changes to SnO and finally sample is completely transformed into $SnO_2$ at higher temperatures. Our X-ray photoemission results showed similar chemical shifts for Sn 3 $d$ core levels for SnO and $SnO_2$ phases while the oxygen 1$s$ level and the valance band spectra showed a remarkable differences pointing to the underlying differences in their electronic structures.